# Carrier-Phonon Scattering Rate and Charge Transport in Spherical and TMV Viruses

*Sanjeev K. Gupta and Prafulla K. Jha*

Department of Physics, Bhavnagar University,
Bhavnagar- 364 022. India

## ABSTRACT

The present paper presents the carrier-acoustic phonon scattering in the spherical and TMV viruses. We demonstrate theoretically that the absorption rate changes in spherical and TMV viruses according to the phonon energy while emission of phonon is limited by the hole energy. The obtained relaxation rate is then used to calculate the conductivity and mobility of viruses. The obtained conductivity for spherical and TMV viruses suggest that the TMV virus is more conducting and therefore may be a good candidate for the connector or wire to be used in the nanoelectronics. The value of resistance obtained for TMV virus is lower than the earlier reported resistance of DNA.

# 1. INTRODUCTION

In the recent years, the fabrication and development of semiconductor devices with nano-dimension has created tremendous interest in integrating the electronics with the biological objects in new and interesting ways [1]. These new integrated devices are known as molecular electronics and are of importance to develop new devices ranging from the medical instruments to other applications. One class of biological object, viruses are found to work as nanotemplates in nanofabrication [2-5]. Virus particles, the smallest living microorganism existing in nature with variety of shapes and sizes are in general comprise a nucleic acid genome that may be either single- or double- stranded DNA or RNA. This is surrounded by a protective coat of protein called a capsid. The virus capsid may be either spherical or helical and is composed of proteins encoded by the viral genome. In helical viruses, the capsid protein binds directly to the viral genome. Spherical virus's capsids completely enclose the viral genome and do not generally bind as tightly to the nucleic acid as helical capsid proteins. These structures can range in size from less than 20 nanometers up to 400 nanometers and are composed of viral protein arranged with icosahedral symmetry [7].

Another issue which has received focus in recent time is the charge transport in these biological objects due to its potential candidature as nanotemplates and connector in the electronic circuit. In addition, electronic excitation and motion of electric charges are well known to play a significant role in wide range of macromolecules of biological objects and is an established fact that these molecules are capable of transporting charges over a distance of at least a few nanometers [8-11]. The rod-shaped viruses, such as tobacco mosaic virus (TMV) and M13 bacteriophage have been utilized as biological templates in the synthesis of semiconductor and metallic nanowires [4,6]. They have been also proposed as elements in the biologically-inspired nanoelectronic circuits and it is expected that the genetically programmed viruses will contribute to the heat generation of nanoelectronic circuits and optoelectronic devices. Recently, it is suggested that the viruses previously thought of only as nanotemplates in the circuits can now actually improve the electronic and thermal properties of the man made nanostructure grown on them because of the suppression of the electron-phonon scattering rate due to the phonon confinement [12]. The availability of different types of viruses, characterized by their size, surface structure, genome sequence and coat protein kinetics expands the range of materials that can be combined in assembly [13]. Tobacco mosaic virus (TMV), Perina nuda viruses (PnV), spherical influenza A viruses, bacillary white spot syndrome viruses (WSSV) and isometric enterovirus 71 (EV71) are the viruses convenient for the applications as nanotemplates to fabricate interconnections or elements of the device structure. While the TMV has a cylindrical rod shape with suitable nanometer scale dimensions of 300 nm long, 18 nm diameter with a 4 nm inner diameter axial channel, the PnV, spherical influenza A, WSSV, EV71 and HIV viruses form the group of spherical viruses normally ranging their size from 20 to 100 nm in diameter.

Though the DNA and viruses have received great attention in the nanotechnology and nanoelectronics, probably there is no report on the transport properties or the carrier-phonon scattering rate for viruses, in contrast to the DNA for which the situation is better [11, 14-22]. In most of these studies performed for DNA, the conductivity and mobility are calculated by considering DNA chain with different shapes such as ropes [14], distorted DNA [15], chemically doped DNA [16] and non-periodic structure [17,18]. Therefore, these do not only make the comparison unjustifiable for the DNA itself but also gives controversial results. Beleznoy et al [11] have calculated the charge transport in DNA stack and compared with the experimentally measured conductivity for similar structure [19]. It should be mentioned that there exists several study [11,14-19], but to the best of our knowledge so far no calculation of charge transport and carrier-phonon relaxation rate even for the DNA chain in one dimensional cylindrical wire and zero dimensional quantum dot forms have been reported. Moreover, there are some controversial reports suggesting different views on the conductivity of DNA right from this being an insulator to the large gap semiconductor [14-19]. Therefore, it is important to investigate the carrier-phonon relaxation time and transport properties for the viruses which are also made up of the proteins. This will not only lead to understand the carrier flow through viruses but may also shed some light on the carrier flow in DNA.

In the present paper, we present a complete calculation of carrier-phonon scattering rate and charge transport in TMV and spherical viruses, which consist a nucleic acid genome that are DNA and of nanometric size. Furthermore, this will be interesting to see not only the effect of confinement of two (wire) and three (dot) dimensional confined systems but also if the emission and absorption scattering rate show some interesting behavior in themselves as well as with the confinement. Of special concern is the problem of transport properties in TMV and spherical viruses in the context of their possible use in nano-electronics. This information may also of help to understand that how energy relaxes in these two types of viruses. The rest of the paper is organized as follows. In section 2 we present the analytical approach to modeling of carrier-phonon scattering rates in TMV and spherical viruses. Results and discussion are presented in section 3 followed by the conclusion in section 4.

## 2. ANALYTICAL AND COMPUTATIONAL APPROACH

In modeling the charge transport in TMV and spherical viruses, while we consider the actual structure of TMV as one dimensional cylindrical wire and spherical viruses as quantum dot or nanoparticles respectively similar to our previous study of vibrational dynamics in spherical virus [23]. These one dimensional cylindrical wire and spherical nanoparticles are made up of lysozyme proteins and therefore all parameters required to model the transport behavior of these molecular structure are considered for the same [24]. This can be justified with the fact that the total acoustic parameter of virus is not available and most viruses are composed of lipids, proteins etc Further, the rigidity of the virus made up of the lysozyme is of the order of 2.7 GPa, which is relatively high. This leads to the assumption that the whole virus can be treated as one isotropic elastic sphere. Moreover, the purpose of the present study is to see the effect of carrier-phonon interactions an approximation to treat viruses as nanoparticles and nano-rods with effective acoustic parameters is justified.

In nanometric spherical and cylindrical TMV viruses, the acoustic modes become discrete due to the size quantization [23, 25]. Their properties can be described by using classical Lamb's model which leads to the low wave number acoustic phonon modes [23, 25] responsible for the dominant contribution to the scattering process. In the present calculation of carrier-acoustic phonon scattering rate and charge transport in TMV and spherical viruses, the usual form of the deformation potential Hamiltonian is considered, $H_{def} = D\nabla \cdot \vec{u}$ [26]. Here $D$ is the deformation potential coupling constant. In these conditions, longitudinal acoustic modes can only couple to the carriers [27] as transverse modes have a null divergence and therefore do not interact with electron via the deformation potential mechanism. The transition probability rate for the carrier to be scattered from an initial state $i$ to a final state $f$, accompanied by the emission or absorption of an acoustical phonon characterized by the frequency $\omega$ can be obtained by using the well known Fermi golden rule. It should be noted that the deformation potential approximation is valid and can be applied in the present study of viruses as the widths of its valence and conduction bands, is much larger than the thermal energy at body temperature (300 K), $K_BT \approx 0.025$ eV [20]. According to the Fermi golden rule the transition rate for the emission/absorption $w^{\{e,a\}}(k,k')$, from an initial carrier-state $|k\rangle$ and an initial phonon state $\left|N_q + \frac{1}{2} \pm \frac{1}{2}\right\rangle$, denoted by $\left|k, N_q + \frac{1}{2} \pm \frac{1}{2}\right\rangle$, to a final carrier-state $\langle k'|$ and a final phonon state $\left\langle N_q + \frac{1}{2} \pm \frac{1}{2}\right|$, denoted by $\left\langle k', N_q + \frac{1}{2} \pm \frac{1}{2}\right|$, per unit time per unit volume is,

$$W^{\{e,a\}}(k,k') = \frac{2\pi}{\hbar} \sum_q \left|M^{\{e,a\}}(q)\right|^2 \delta(E(k') - E(k) \pm \hbar\omega) \tag{1}$$

Here $M^{\{e,a\}}(q)$ is the matrix-element considers the deformation potential due to electron-acoustic phonon interaction and can be expressed as [28]

$$M^{\{e,a\}}(q) = \left\langle k', N_q + \frac{1}{2} \pm \frac{1}{2} \left| H^{dp}_{e-ph} \right| k, N_q + \frac{1}{2} \pm \frac{1}{2}\right\rangle \tag{2}$$

Where, $H^{dp}_{e-ph} = \sum_q C(q)\left[a_q e^{iq.r} + a_q^\dagger e^{-iq.r}\right] \tag{3}$

and, $|C(q)|^2 = \hbar D^2 q / 2\rho V_l \tag{4}$

The carrier-acoustic phonon scattering rate for an isotropic system can be obtained using deformation potential interaction for non-degenerate carriers at $\Gamma$-point. Here $\rho$ and $V_l$ are density and longitudinal velocity of sound in virus. $N_q = 1 / \left(e^{\hbar\omega/K_B T} - 1\right)$ is the phonon occupation number. In this limit, the carriers are assumed to be confined in the spherical or cylindrical virus by an infinite potential at its surface. Thus the initial state can be given as [28]

$$|k\rangle = \frac{e^{ik.r}}{\sqrt{V}} |\varphi(r,q)\rangle \tag{5}$$

Where, $\varphi(r,q)$ is the wave-function and can be expressed for spherical and cylindrical TMV virus as [28]

$$j_{nlm}(r,q,f) = j_l(x_{nl} r/a) Y_l^m(q,f) \quad \text{[for spherical virus]} \tag{6}$$

and,
$$j(r,q) = \left\{ \frac{1}{\sqrt{p}} Y_0^0 J_0\left(\frac{X_1^0}{a} r\right) \right\} \quad \text{[for TMV virus]} \tag{7}$$

Where, $n=1, 2, 3\ldots$ is the radial quantum number and $l$ & $m$ are angular quantum number, $Y_l^m$ is spherical harmonics and $j_l(x)$ is the Spherical Bessel's function, $x_{nl}$ is nth zero of $j_l(x)$ and $V$ is the volume. The ground state for spherical virus are (1, 0, 0) and for the threefold degenerate first-excited state $(n, l)$ equal to (1, 1) and $m=0$. For TMV virus $x_1^0$ being the position of the first zero of $j_0(x)$ and $Y_1^0$ being equal to $j_1(x_1^0)$. By using Eq. (6) in the solution of matrix element (Eq. (2)), the carrier-phonon scattering rate for the emission for viruses can be written as,

$$W^{\{e,a\}}(k,k') = \frac{1}{t^{\{e,a\}}(k,k')} = \frac{0.727 p (N_q + 1) m^5 V_l^3 D^2 a^2}{r \hbar^6} \quad \text{[for spherical viruses]} \tag{8}$$

and,
$$\frac{1}{t^{\{e,a\}}(k)} = \sum_n \frac{p}{|Y_1^0|^4 \hbar^2} \frac{mD^2}{rp^2} \cdot \frac{w^3 |b_n(\tilde{q})|^2}{c_l^4 |S_n(\tilde{q})|} \cdot \frac{|F_n(\tilde{q}_l p)|^2 (N_{q+} \tfrac{1}{2} + \tfrac{e}{2})}{(q+k) - \left(\tfrac{em}{\hbar}\right) dw/dq} \quad \text{[for TMV viruses]} \tag{9}$$

The carrier-phonon scattering rate for the absorption for viruses can also be calculated by just replacing the $(N_q+1)$ by $N_q$ in above equation.

The low-field acoustic-phonon-limited carrier $\mu$ mobility is calculated in the relaxation time approximation as

$$s = \frac{ne^2}{m^*} \langle t \rangle \tag{10}$$

$$m = \frac{e}{m^*} \langle t \rangle \tag{11}$$

Where, τ is the carrier-phonon scattering rate and *n* is the concentration of charge carrier which can be expressed as,

$$n = \oint f_0(k) N(k) dk \tag{12}$$

Where, $N(k)$ =Density of states (DOS) and $f_0(k) = \exp\left(\dfrac{(e(k) - e_F)}{k_B T}\right)$, with $e_F$ as Fermi energy and, $e(k) = \dfrac{\hbar^2 x_{nl}^2}{2ma^2}$. The density of states $N(k)$ in spherical and TMV viruses are given as,

$$N(E) = \frac{dN}{dE} = 2\left[\frac{1}{V}\right] d(E - E_n) \quad \text{[for spherical virus]} \tag{13}$$

and, $$N(E) = \frac{1}{\pi \hbar} = \sum_n \sqrt{\frac{m}{2(E - E_n)}} s(E - E_n) \quad \text{[for TMV virus]} \tag{14}$$

Where, $E_{bottom}$ = lowest quantized state of the spherical virus system and $E_n$= energies of the quantized states of the TMV viruses. The present calculation considers the $\epsilon_F$ =0.265eV. Since Fermi energy for viruses is not available. We have considered this value as an approximation from the guanine's [20]. A hole placed in a guanine's HOMO is only about 0.2 eV above the next lower occupied orbitals and can begin to migrate through DNA to find other easily oxidizable site, like other guanine's or sequence of guanine's.

3. **RESULTS AND DISCUSSION**

In the following, we present results on the carrier-phonon relaxation rate and transport properties such as mobility and conductivity for the spherical and TMV viruses by using Eqn. (6-11). As far as the sound velocities for considered viruses are concern, we used the values of velocities to estimate the carrier-phonon relaxation rate and transport as of protein crystals as the acoustic parameter of a total virus is rarely studied. The deformation potential

D≈6.0 eV/nm obtained from the expression given in ref [11] is used in present calculation. The parameters required obtaining deformation potential are used for the lysozyme [24]. The quantity D≈6.0 eV/nm is also used by Conwell and Basko in their study on the hole traps in DNA [29]. This value is also fairly close to the ionization potential ~6-8 eV DNA bases in vacuum [30]. In the present study the emission and absorption rates of carrier-acoustic phonon have been calculated for both carriers electron and hole, however, only for hole have been reported in the present paper as the trend is similar. Moreover, it is observed that the charge migration through DNA have been centered on hole transfer due to its widely varying distance dependence [9]. It is interesting to see in what follows that the absorption rate changes significantly for the phonon energy in comparison to the emission rate for both kind of viruses considered in the present study.

Figures 1-4 present the phonon emission and absorption rates for the TMV and spherical viruses for different phonon energy $E_{phonon}$. These figures, however, show the plots for higher $E_{phonon}$ but the trend is similar for lower $E_{phonon}$ obtained form the low frequency acoustic modes for both kind of viruses [23, 25]. The higher phonon energies chosen are not related to the actual virus's structure but only for the simulation purpose. However, the lower phonon can be related to the actual virus's structure. It is seen from these figures that there is no emission of phonon before the $E_{phonon}$ i.e. this is the cutoff energy for the phonon emission. This suggests that the rapid relaxation of carrier is possible only in limited ranges of $E_{phonon}$. To have a better visualization and comparison, the phonon emission and absorption rates for both spherical and cylindrical (TMV) viruses are presented in Table I. The Table I is clearly able to bring out the difference in different scattering rates in different geometries and therefore able to show the effect of dimensional confinement on the phonon emission and absorption rates. These figures along with the Table I reveal that the absorption and emission

rates both decreases going from the TMV (wire) to the spherical (QD) and emission rates are in the range of only few picoseconds. However, a remarkable feature is observed from table I that the absorption rate changes significantly according to $E_{phonon}$, while the emission is limited by the hole energy equivalent to the phonon energy. This may be seen as the existence of several of phonons present near a particular phonon energy [31] and therefore absorption of many or multi phonons are expected [32]. Therefore, if multiple absorption of phonon is considered the two rates may become comparable to the energy emission by phonon. This suggests that the carriers available for the conduction can cross the energy valley equivalent to the phonon energy easily and excited.

It is a well known fact that the motion of charge carriers play a significant role in a wide range of macromolecules of biological interest and recent interest in viruses, charge transport has been spurred by its relevance to potential applications in molecular electronics [33, 34]. However, of special concern is the problem of mobility and conductivity of charge carriers in viruses in the context of possible use of the viruses as wires and templates in nanoelectronics. To calculate the conductivity, the phonon energies are used from the recently reported low frequency from ref [23] and ref [25] for spherical and TMV viruses respectively. The temperature is considered room temperature i.e. 300 K and frequency of lowest phonon modes in both the cases. The calculated conductivity at room temperature for the 20 nm spherical virus is $8.73 \times 10^{-2}$ $\Omega^{-1}cm^{-1}$ and $1.68$ $\Omega^{-1}cm^{-1}$ for electron and hole respectively. The mobility is $0.35 \times 10^{-5}$ $cm^2/V.S$ and $6.83 \times 10^{-5}$ $cm^2/V.S$ for electron and hole respectively. These quantities are significantly different for the TMV virus (one dimensional cylindrical wire) for both electron and hole. While the conductivity is $3.91$ $\Omega^{-1}cm^{-1}$ and $11.03$ $\Omega^{-1}cm^{-1}$ for electron and hole respectively, the mobility comes out $1.71 \times 10^{-6}$ $cm^2/V.s$ and $0.42 \times 10^{-6}$ $cm^2/V.s$ for electron and hole respectively for the TMV virus of 20 nm diameter and length

300 nm. This clearly demonstrates the change in two structures obviously arising due to the confinement effect on electron-phonon interaction, density of states and carrier concentration. It is seen that the conductivity decreases going down to dimensional confinement i.e. from TMV to spherical viruses. This looks at first instance due to the decrease in the carrier concentration in spherical virus. However, it is not possible to comment on the values of mobilities and conductivities obtained in the present study but gives confidence in these numbers as they are close to the value obtained by ac conductivity measurement for the DNA [11] an another class of biological object and is potential candidate to be used in molecular electronics. The present calculated results on conductivity and mobility [34, 35] are in the order of reported on DNA with different structures, but with higher values particularly for TMV virus [36, 37]. It is to be noted that the present calculation is first attempt to investigate the transport properties for spherical and TMV viruses. Further the present study considers the protein made up of lysozyme instead of E. coli. The high value of conductivity for TMV virus shows its utility as the medium for the carrier transport. For TMV viruses considering its diameter of $r$=20 nm and cross-section A= $\pi r^2$, the conductivity of 11.03 (Ohm-cm)$^{-1}$ and length 300 nm the resistance R ≈ 0.2x10$^6$ Ohms for hole carrier.

The present value of R ≈10$^6$ Ohms for TMV virus shows that the TMV viruses are more conducting than the DNA, which has been reported as insulating due to its high resistance R ≈10$^8$ [34] and 10$^{10}$ [35] Ohms observed from the two different studies. Moreover, the present study also suggests that the TMV virus which is an one dimensional cylindrical wire form can be the better choice than spherical virus and DNA one dimensional chain to work as a conductor and thereby can be utilized as inter connector or wire with man made nanostructures i.e. semiconductor nanostructures. The reason for this behavior of TMV virus is not clear, however, it may be due to some kind of phonon assisted polaron hopping [38] or due to the reduction in band gap formed due to inherent one dimensionality nature [39].

## 4. CONCLUSIONS

Finally, in the present paper, the relaxation time via carrier- acoustic phonon scattering and charge transport for TMV and spherical viruses have been reported. While the absorption rate changes significantly the emission rate is limited by the hole energy. The emission rates are in the range of few picoseconds. This information suggests existing of several of phonons about a particular phonon energy. The present study is able to bring out the nature of relaxation times or how energy relaxes in two different types of viruses and hence to understand the transport behavior. The conductivity is obtained reasonably high for the TMV virus, opposite to the general observation of DNA chain which has been reported either semiconducting or insulating, For the difference in the values of conductivity, it is important to recognize that the conductivity is extremely sensitive to the type of proteins, virus or DNA extractions and origins in addition to the dimensionality of charge transport. It is hoped that the present study may be of help to correlate the structure of biomolecules to their mechanical properties.

**ACKNOWLEDGMENT**

This work was supported by the DAE-BRNS, Project No. 2005/37/6/BRNS/791.

**Table Captions**

Table I.: The Emission and Absorption rates for TMV and Spherical Viruses. The rates are in 1/Second.

**Figure Captions**

Figure 1. Variation of Carrier-Phonon Emission Rates with Energy of Carriers for TMV virus.

Figure 2. Same as fig. 1 but for Absorption rates.

Figure 3. Same as fig. 1 but for Spherical Virus.

Figure 4. Same as fig. 3 but for Absorption rate.

Table I.

| $E_{phonon}$ | Spherical Viruses | | TMV Viruses | |
|---|---|---|---|---|
| | Absorption rate | Emission rate | Absorption rate | Emission rate |
| 50 meV | $10^8 \sim 10^{11}$ | $10^{10} \sim 10^{11}$ | $10^9 \sim 10^{13}$ | $10^{12} \sim 10^{13}$ |
| 100 meV | $10^5 \sim 10^{11}$ | $10^{10} \sim 10^{11}$ | $10^6 \sim 10^{14}$ | $10^{12} \sim 10^{13}$ |
| 150 meV | $10^2 \sim 10^{11}$ | $10^{10} \sim 10^{11}$ | $10^3 \sim 10^{14}$ | $10^{12} \sim 10^{13}$ |
| 250 meV | $10^{-3} \sim 10^{11}$ | $10^{10} \sim 10^{11}$ | $10^{-4} \sim 10^{14}$ | $10^{12} \sim 10^{13}$ |

Figure 1.

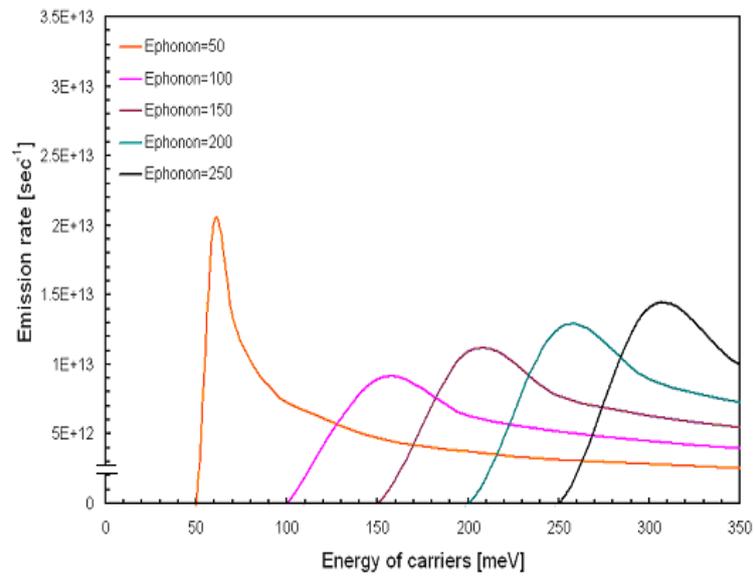

Figure 2.

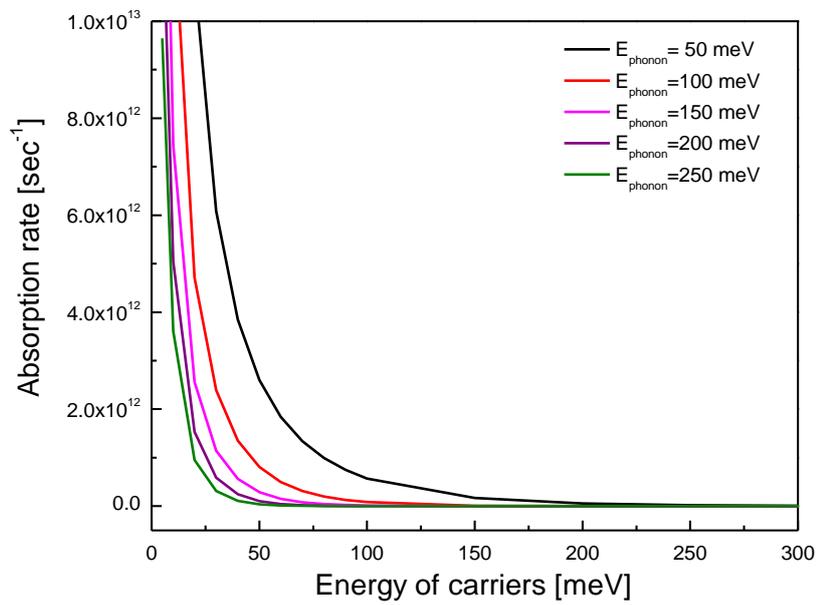

Figure 3.

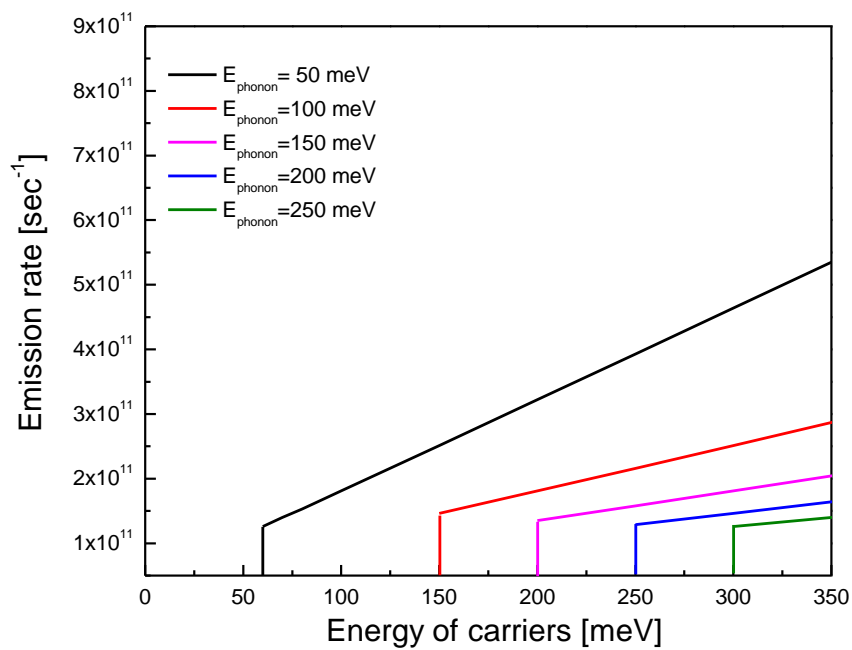

Figure 4.

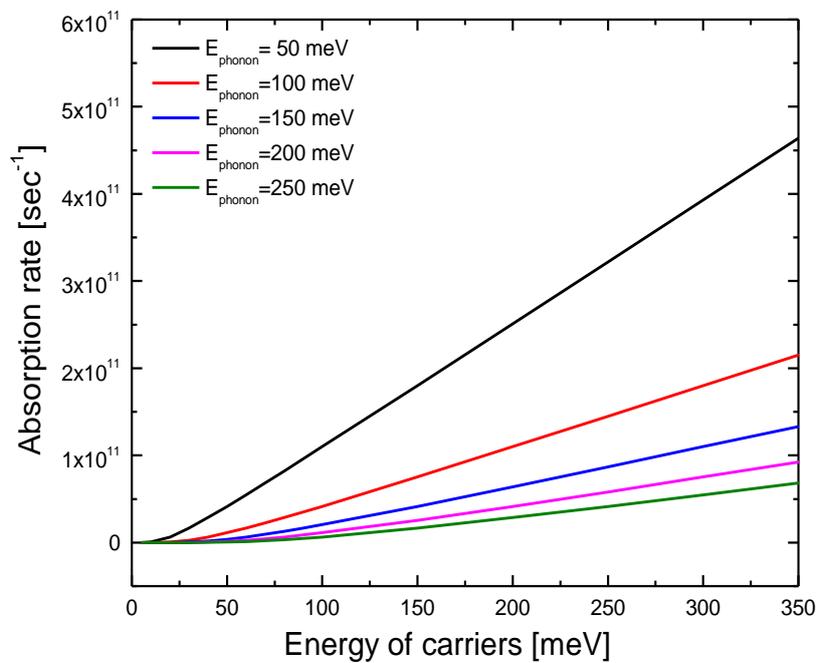